# Edge-enabled Optimized Network Slicing in Large Scale Networks

José Jurandir Alves Esteves*†, Amina Boubendir*, Fabice Guillemin* and Pierre Sens†
*Orange Labs, France
†Sorbonne Université / CNRS / Inria, LIP6, France
{josejurandir.alvesesteves, amina.boubendir, fabrice.guillemin} @orange.com, pierre.sens@lip6.fr

*Abstract*—In this demo paper, we consider the network slice placement optimization problem and give some insights into a fast heuristic algorithm tailored to placement in large scale networks. We consider an online optimization scenario with multiple and volatile network slice request arrivals and we showcase the applicability of the proposed Edge-enabled network slice placement solution through a Proof-of-Concept illustrated by large scale networks scenarios.

*Index Terms*—Network Slicing, Optimization, NFV, Large Scale Networks, Heuristics, Placement.

## I. Context and Motivation

Network Function Virtualization (NFV) is transforming the architecture and the operation of telecommunications networks by breaking the link between functions and their hosting hardware. The physical network infrastructure then becomes a pool of resources shared by various virtualized network functions (VNFs), which are themselves the components of virtual networks. The life-cycle of VNFs can be managed separately from that of the underlying physical infrastructure.

An NFV-based network can be viewed as a programmable platform that enables the deployment of VNF chains forming logical networks over the same shared Physical Substrate Network (PSN). This has given rise to the concept of Network Slicing, which takes benefit of the logical and/or physical separation of network resources to allow multi-tenancy support, customization and isolation of Network Slices [1].

Network slice placement appears as an optimization problem that consists of choosing the servers of the PSN in which the VNFs composing a Network Slice can be deployed and which physical links to use in order to steer traffic between these VNFs. This is an optimization problem with objectives (e.g., minimizing resource consumption, optimizing a specific QoS metric, etc.) that must be satisfied [2], [3].

Numerous papers about network slice placement and its variants use heuristic-based approaches to solve associated optimization problems. However, addressing scenarios with large scale PSNs and Network Slice Placement Requests (NSPR) or the Edge-specific constraints and thus the direct impact on QoS metrics (notably, E2E latency) remains a challenge. This paper describes the Proof-of-Concept (PoC) and a corresponding demonstration of the network slice placement problem considered in a detailed paper [4].

We give hereafter an insight into our Edge-enabled approach to dealing with network slice placement. The proposed demo specifically addresses an Edge-enabled Network Slicing in large scale network scenario. Our solution relies on a Network Slice Placement algorithm that is adapted to large scale network scenarios and integrates both edge-specific constraints related to user location and strict end-to-end (E2E) latency requirements.

The paper is organized as follows. First, we highlight the objectives in Section II-A. Then, we present the solution architecture in Section II-B and detail its implementation in Section II-C. The planned demonstration is described in Section III. We conclude the paper in Section IV.

## II. Large Scale Network Slice Placement Optimization: Proof-of-concept Architecture

### A. Objective and Overview

The objective of this PoC is to show the applicability of the proposed Network Slice Placement algorithm presented in the associate full paper [4]. We precisely analyse the capacity of this solution of providing accurate placement solutions with small execution time in an online optimization environment, where NSPRs are not known in advance and their demands are volatile since they stay in the PSN for a limited duration.

We evaluate two different placement algorithms: 1) an ILP-based algorithm that deterministically calculates the placement decisions and 2) a heuristic called Power of Two Choices (P2C), that employs a simpler but efficient policy that quickly generates placement solutions more adapted to large scale network scenarios. Both algorithms are designed to optimize the total resource consumption under constraints on the PSN capacities (CPU/RAM hosting nodes capacities, bandwidth and latency capacities on the links) and NSPR requirements (CPU/RAM VNFs requirements, bandwidth and latency requirements on the virtual links).

### B. Proof-of-Concept Architecture Description

The architecture of our Network Slice Placement scheme is given in the Figure 1. The NSPR generator is used to generate NSPR arrivals. The PSN database stores the data about the available resources of the PSN. NSPR requirements and PSN available resources data are used as input by the Placement module. This latter implements the ILP-based and P2C algorithms.

Both algorithms calculate: i) a VNF placement decision, that is, where each VNF of the NSPR is to be placed and ii) a VNF

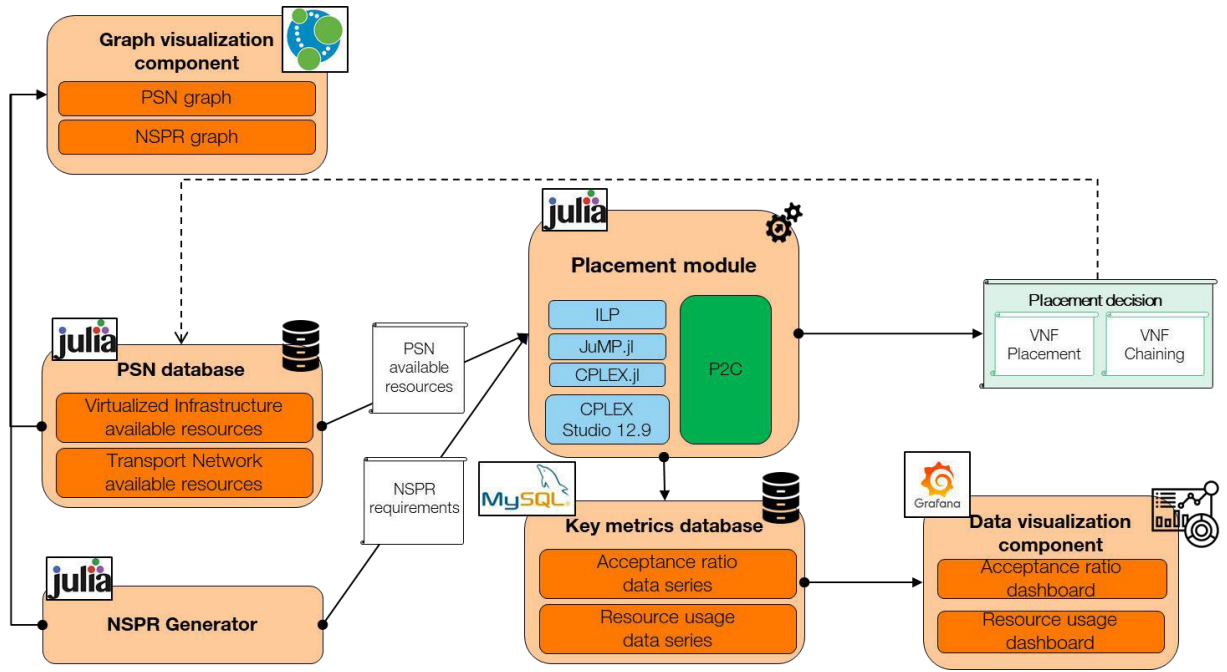

Fig. 1. Proof-of-Concept Architecture: Optimized Network Slice Placement Supporting Edge and Large Scale Networks

chaining decision, that is, which paths in the network to use to interconnect the different VNFs. The Placement module can be configured to use one of the Placement algorithms or both if comparison of Placement solutions is necessary. We compare the performance of both algorithms in our demonstration.

Once the calculation of the Placement decision is done for one NSPR, an update of the available resources on the PSN is made and some key performance metrics are registered in form of data series the Key metrics database: the acceptance ratio of network slices and the resource usage. These time series are used by the Data visualization component to build two dashboards: an acceptance ratio dashboard and a resource usage dashboard. Both dashboards are used to show the performance of the algorithms in real time. Finally, the graph visualization component is used to allow the visualisations of the PSN and NSPR graphs.

### C. Solution Implementation Description & Used Tools

We have implemented the proposed Network Slice Placement solution and we describe below the different tools used to implement the different components of our solution:

- Julia: The [5] version 1.1 is used to implement different elements of our solution:
  1) the PSN database, that is actually represented by a series of data frames loaded from csv files,
  2) the NSPR generator that is actually a function that receives some parameters representing the CPU, RAM, bandwidth and latency requirements of the NSPRs to be generated,
  3) the placement algorithms.

The ILP-based Placement algorithm is developed using the JuMP.jl package that allows the development of a generic ILP formulation that can be solved with any linear solver. The CPLEX.jl package is an interface to the CPLEX solver used to solve the ILP. The classes and functions used to implement the P2C heuristic were designed and implemented from scratch using only default Julia packages.

- CPLEX: The default branch-and-bound algorithm from *ILOG* CPLEX [6] in its 12.9 version is used to solve the ILP formulation.
- Neo4j: A Neo4j graph database represents and and displays the PSN graph and the NSPR graph and its requirements.
- MySQL: We use the MySQL database manager system to implement the Key metrics database with one table for the Acceptance ratio data series and another one for the Resource usage data series.
- Grafana: We use the Grafana tool to implement the Data visualization component in which we represent two dashboards using Key metrics MySQL database as datasource.

### III. PLANNED DEMONSTRATION

We perform an emulation of the network slice placement decision making process using our implementation.

### A. Considered Demonstration Scenarios

We emulate a realistic PSN topology with 21 data centers (DCs) with different resource capacities.

We consider 15 Edge Data Centers (EDCs) as local DCs with small resources capacities, 5 Core Data Centers (CDCs)

as regional DCs with medium resource capacities, and 1 Central Cloud Platform (CCP) as a national DC with big resource capacities. We consider 1008 hosting nodes distributed among these DCs offering IT resources to support the VNFs. See Figure 2.

The proposed emulation is meant to demonstrate the deployment of optimized network slice placement decisions calculated with the P2C heuristic and to compare it with placement decisions calculated by the ILP-based algorithm. PSN and NSPR are displayed by using Neo4j in Figures 2 and 3, respectively.

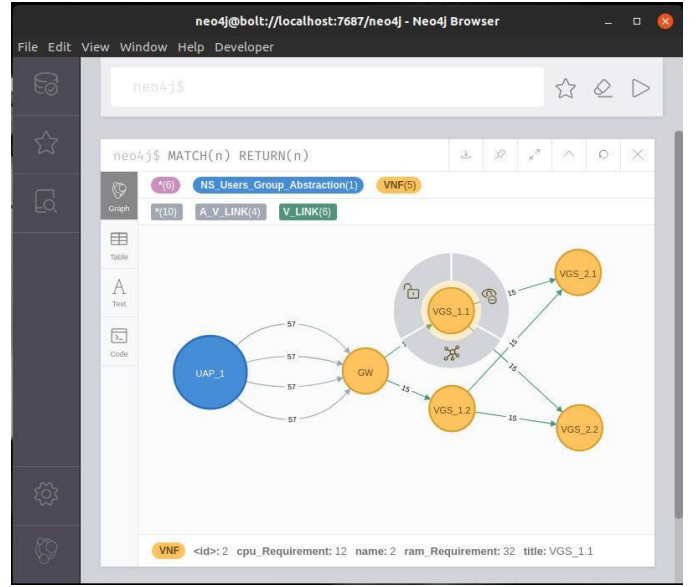

Fig. 3. Network Slice Placement Request view using the Graph visualization component

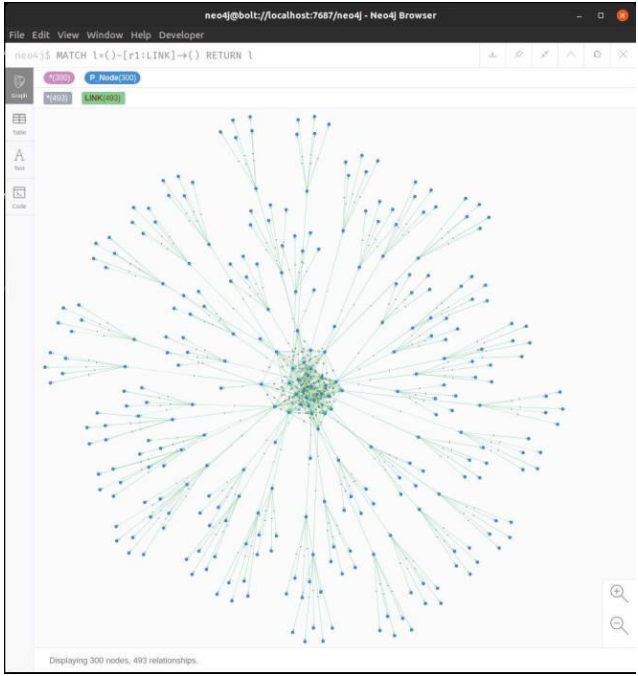

Fig. 2. Physical Substrate Network view using the Graph-based Visualization Component

### B. Demonstrated Aspects

We show four different aspects in our demonstration:
1) the capacity of both methods of achieving different network slice requirements (CPU, RAM, end-to-end latency),
2) the capability of both algorithms of sustaining a high network slice acceptance ratio when we have a critical network load,
3) how a variation of the amount of resources in each hosting node can impact the performance of the algorithms,
4) how the variation of size of the NSPRs in terms of the number of VNFs can impact the performance of the algorithms.

## IV. CONCLUSION

This paper presents a PoC of a Network Slice Placement algorithm proposed to support Edge-empowered and large scale networks.

We have described the implementation of the proposed solution in order to emphasize its feasibility and its accuracy when considering network Slice Placement over Edge Data centers and large scale networks scenarios.

The proposed visualization components based on Neo4j Graphs and Grafana are of utmost importance as they allow a fast analysis and understanding of the algorithms performance metrics. We show how it can provide low execution times in large scale network scenarios. Our approach is generic and can be applied to different Network Slice use cases.

ACKNOWLEDGMENT

This work falls within the framework of 5GPPP MON-B5G collaborative project: www.monb5g.eu.

REFERENCES

[1] I. Afolabi, T. Taleb, K. Samdanis, A. Ksentini, and H. Flinck, "Network slicing and softwarization: A survey on principles, enabling technologies, and solutions," *IEEE Commun. Surv. Tut.*, vol. 20, no. 3, pp. 2429–2453, 2018.
[2] M. Mechtri, C. Ghribi, and D. Zeghlache, "VNF placement and chaining in distributed cloud," in *Proc. 9th IEEE Int. Conf. Cloud Comput.*, 2016, pp. 376–383.
[3] J. G. Herrera and J. F. Botero, "Resource allocation in NFV: A comprehensive survey," *IEEE Trans. Netw. Service Manage.*, vol. 13, no. 3, pp. 518–532, 2016.
[4] J. J. Alves Esteves, A. Boubendir, F. Guillemin, and P. Sens, "Heuristic for edge-enabled network slicing optimization using the power of two choices, accepted," in *16th International Conference on Network and Service Management (CNSM)*, 2020.
[5] J. Bezanson, S. Karpinski, V. B. Shah, and A. Edelman, "Julia: A fast dynamic language for technical computing," *arXiv preprint arXiv:1209.5145*, 2012.
[6] I. ILOG, "CPLEX optimization studio," 2014.